\documentclass[12pt]{article}
\usepackage{sao1}
\usepackage{amssymb}

\begin{document}
\begin {center}
{\large\bf
Gravitational collapse as the source of gamma-ray bursts
\footnote{to be published in Proceedings of the Conference
``Problems of Practical Cosmology",
see http://ppc08.astro.spbu.ru/index.html}
}\\
\vspace{4mm}
{V.V.Sokolov}\\
\vspace{4mm}
{$^.$ Special Astrophysical Observatory of RAS, Russia\\
}
{\it May 21, 2008}
\end{center}
\vspace{-7mm}

\def\g{$\gamma$}
\begin{abstract}
If the threshold for $e^{-}e^{+}$ pair production depends
on an angle between photon momenta, and
if the $\gamma$-rays are collimated right {\it in} gamma-ray burst (GRB)
source then another solution of the compactness problem is possible.
The list of basic assumptions of the scenario describing the GRB
with energy release $< 10^{49}$\,erg is adduced:
the matter is about an alternative to the ultrarelativistic fireball
if {\it all} long-duration GRBs are physically connected
with
core-collapse supernovae (SNe).
The questions about radiation pressure and how the jet arises on account of
even a small radiation field asymmetry in a compact GRB source
of size $\lesssim 10^8$\,cm,
and observational consequences of the compact model of GRBs are considered.
\end{abstract}

\section{Introduction: the root of the problem}

   Gamma-ray bursts
(GRBs) are the brief ($\sim$\,0.01-100\,s), intense flashes of $\gamma$-rays
(mostly sub-MeV) with enormous electromagnetic energy release up to
$\sim 10^{51} - 10^{53}$\,erg.
The rapid temporal variability, $\delta T \lesssim 10$\,msec,
observed in GRBs implies {\it compact} sources with a size smaller
than $c\,\delta T \lesssim 3000$\,km.

   But a problem immediately arises
for distant GRB sources (e.g. [1, 2]):
%Carrigan and Katz 1992
too large energy ($>10^{51}$\,ergs) is released in the observed
(for the most GRBs) soft
$\gamma$-rays ($<511$ keV and up to 1 MeV) in such a  small volume
for the sources at cosmological distances ($>1$\,Gpc).
For a photon number density
$n_{\gamma} \sim (10^{51}$erg$/(m_e c^2))/(c\,\delta T)^3
\sim 10^{57}/(3000\,$km$)^3 \sim 10^{32}$cm$^{-3}$ two
$\gamma$-ray photons with a {\it sum energy} larger than $2 m_e c^2$
could interact with each other and produce electron positron
pairs. The optical depth for pair creation is given approximately
by $\tau_{e^+e^-}\sim  n_{\gamma} r_{e}^2 (c\,\delta T) \sim
10^{16}$, where $r_e$ is the classical electron radius
$e^2/(m_ec^2)$
(the cross-section for pair production is $\sim r_{e}^2$ or
 $\sim 10^{-25}$cm$^2$
at these semirelativistic energies).
It is the essence of
a so-called ``compactness problem":
the optical depth of the relatively low energy photons
($\sim 511$\,keV)
must be so large that these photons could not be observed.

  In the popular ultrarelativistic fireball GRB model [3, 4]
%(Piran 1996; 1999; 1999a; 2004)
in this definition a role of
the high-energy photons is emphasized:
the $\gamma$-ray photons with energies much larger than
$m_e c^2$ (or $\gg 1$\,MeV) could interact with lower energy ($< 511$\,keV)
``target'' photons
and produce $e^{-}e^{+}$ pairs.
(e.g. [4]).
%(e.g. Piran 1999).
 In the ultrarelativistic fireball model it is supposed that 
the ``heavy"/hard (or high-energy) photons {\it must be present}
in all GRB spectra as high energy tails which contain a
significant amount of energy.
So, (see e.g. [5]) the optical depth
{\it of the high-energy photons} ($\gg 1$\,MeV)
would be so large that these photons could not be observed.
In this theory the size of the region where the
GRB prompt emission arises
must be $\sim 10^{15} - 10^{17}$\,cm [6],
%(Beloborodov 2004),
if it is supposed that radiation (with 100\,MeV and 10\,GeV photons) is 
generated by ultrarelativistic jets moving with huge Lorentz factors 
$\sim $\,100-1000.

Below we will try to understand the {\it observational soft}
(in the meaning of photon energies) GRB spectrum
in a compact GRB model implies the GRB source with the size
of $c\,\delta T \lesssim 10^8$\,cm,
e. d. without involving huge kinematical motions
of the radiating plasma, or without so enormous Lorentz factors.
%$\Gamma \gg 1$.
It concerns with another attempt of  solving the compactness problem,
namely, the dependence of the threshold for $e^{-}e^{+}$ pair production
on the angle between photon momenta,
a photon collimation {\it in} the source
and the dependence of this collimation on GRB photon energy
must be accounted for.
Taking into consideration the compactness of the source and the fact 
that all long-duration GRBs are physically connected with
core-collapse supernovae (SNe), it may be supposed that when observing 
these GRBs we directly observe the gravitational collapse of a massive 
and compact star core.

\section{Typical GRB spectra and typical photon energies}

The GRB spectra are described in a review by Fishman and Meegan [7],
%Fishman and Meegan (1995),
see also the catalogue of the spectra by Preece et al. [8].
Typical observational GRB spectra turned out to be very diverse,
but yet these are mainly soft (but not hard) $\gamma$-ray quanta.
It has been known  since the moment of GRBs discovery,
when their spectra were presented in energy units:
e.g., see a review by Mazets and Golenetsky [9],
and many authors [10-14] point to the same again.
%(Lamb et al. 2003; Baring \& Braby 2004;
%Liang et al. 2004; Atkins et al. 2003; Gialis \& Pelletier 2004).
Almost all GRBs have been detected in the energy range
between 20 keV and 1 MeV.
{\it A few} $\gamma$-ray quanta have been observed in GRBs above 100 MeV.
In a review by Piran [3] also has paid attention
to a puzzle of the origin of narrow
distribution for the typical energy
of the observed GRB radiation ($E_p < 511$\,keV, [8]).
More, by 2000 it was clear that there were other two GRB classes:
X-Ray Flashes (XRF) and X-Ray Rich Gamma Ray Bursts (XRR GRB) [15, 16].
%(Heise et al. 2001; Amati et al. 2002).
These are GRBs either {\it without} (XRFs) or almost without (XRR GRB)
$\gamma$-ray quanta.

     Thus, for usual (mostly sub-MeV) GRBs,
still there are too many lower energy $\gamma$-ray
photons in a small volume $R^3$ with $R\sim c\,\delta T \lesssim 3000$ km.
The observed fluxes give an estimate
of a total GRB energy release to be of $\sim 10^{51}$ ergs
in the form of
just these {\it low energy} photons,
 or this ``standard"\, estimation ($\sim 10^{51}$\,erg)
was obtained from
typical observational GRB spectra of just these,
most frequently observed low-energy photons with
the {\it semirelativistic} energies, up to 1 MeV, basically.
(It is natural that the photon
density was estimated using the simple assumption of spherical symmetry ---
see below.)

   Nevertheless, if these
 theoretical (rather than observational) statements [17]
%(Piran 1996)
 on the possibility that {\it as though} (1) all GRB spectra have
 high energy tails and (2)
 the observed GRB spectra are non-thermal,
are true indeed,
the fireball theory [4, 18]
%(Piran 1999, 1999a)
with huge Lorentz factors
is the only possible theoretical alternative for GRBs.
It should be admitted though that
the standard optically thin synchrotron shock emission model
explains everything, except the observational spectra of GRBs themselves [19].
%(Preece et al. 2002).
But for all that, it was left out of account that these (``target'') photons
with $E_p < 511$\,keV are just the observed typical GRBs.
So, it turns out that the main task, according to the standard fireball model,
is not the
explanation of this observed soft GRB spectrum in terms of photons'
energy (frequency), but the investigation of rare cases of  release
of hard quanta with energy of more than or $\sim 1$\,GeV.

   As a result, the origin of the {\it observed} and substantially soft GRB
spectra with a big number of photons
%up to
$\lesssim 1$\,MeV remains
not properly understood. It is especially incomprehensible against
the background of conjurations about the huge gamma factor that is supposed
to solve the compactness problem. But the question remains: why are
mainly soft GRB spectra  observed at ultrarelativistic motions
of radiating plasma supposed in the fireball model? And what is more,
as was noted above,
sometimes the GRB spectra do not contain \g-ray quanta at
all, as, for example, XRFs known already before 2000 [20].
%(Heise et al. 2001).
Thus, when solving the compactness problem, we somehow
imperceptibly incurred another problem of strong contradiction
between the ultrarelativistic Lorentz factor $\Gamma \sim$ 100-1000
(in the fireball model with 100\,MeV and 10\,GeV photons)
and observed soft ($\sim$ or $<1$\,MeV) \g-ray (GRB,\, XRR GRB)
and X-ray (XRF) radiation of the most classical GRBs.
Moreover, it is also important to point out here
that the observed {\it black-body} prompt GRB radiation with a
temperature $kT\sim 100$\,keV [21, 22]
%(Ghirlanda et al. 2003; Ryde 2004)
is inconsistent with the Lorentz factor $\approx 10^2 - 10^4$ for the
reason that the mean observed temperature can easily exceed $kT$=1\,MeV in
cosmological fireballs [23].
%(Piran and Shemi 1993).

\section{The threshold for $e^{-}e^{+}$ pair production}

The compactness problem was mentioned
(before 1992, i.e. before the BATSE/EGRET mission)
in connection with the famous burst of 1979 March 5 in the
Large Magellanic Cloud. Already then a possibility of
a {\it photon collimation right in the source}
for explanation of observed soft spectra
was not excluded [1]
%(Aharonian and Ozernoy 1979)
because the cross-section of electron-positron pair production
$\sigma_{e^-e^+}$ (and annihilation also) depends not only on energy,
but on the angle between momenta of colliding particles.
For the first time
in the paper by Aharonian and Ozernoy [1], and than later
by Carrigan and Katz [2],
a lot of interesting was said in connection with collimation
of $\gamma$-rays leaving GRB source with high photon density in it.

It seems that just the collimation in GRB source solves
the notorious compactness problem indeed.
The paper by Carrigan and Katz [2]
tells about modeling the observed GRB spectra allowing for the
electron-positron pair production effects.
These effects could produce effective collimation of the flux
because of kinematics of the two-photon pair production:
the opacity ($\tau_{e^-e^+}$) is also a sensitive
function of the {\it angular} and {\it spectral} distribution of the
radiation field {\it in the source}.

The argument proceeds as follows:
{\it two} photons with energies $E_1$ and $E_2$,
which are above the threshold
energy ($E_1+E_2 > 2\cdot E_{th}$, $E_{th}= \sqrt {E_1 E_2}$)
for electron-positron pair production
\begin{equation}
E_{th}^2 = E_1 \cdot E_2 \geq 2(m_e c^2)^2 / (1-cos\theta_{12})
\end{equation}
may produce a pair,
 where $2(m_e c^2)^2 = 2(511\,$keV$)^2$
and $\theta_{12}$ is the angle between the directions
of the two $\gamma$-rays.
The cross section for pair production reaches
the maximum at a finite center-of-momentum photon energy:
e.g. $E_1 + E_2 > 2\cdot E_{th} = 2\cdot$511\,keV
for $\theta_{12} = 180\degr$,
or $E_1 + E_2 > 2\cdot E_{th}\approx 2\cdot$700\,keV
for $\theta_{12}\approx 90\degr$),
or $E_1 + E_2 > 2\cdot E_{th}$ going to infinity ($\gg$1\,MeV)
for $\theta_{12}\approx 0\degr$.

  If the source photon spectrum is not sharply peaked,
the relatively high-energy photons ($E > E_{th}$) will,
therefore, form pairs predominantly with relatively
low-energy photons ($E < E_{th}$).
It means that the observed (or the emergent) GRB spectra will be soft,
since the high-energy photons will be held by the threshold
of pair production.
Thus, because any {\it reasonable} source spectrum will contain much more
low- or  moderate-energy photons ($\lesssim 511$\,keV)
than high-energy photons,
the emergent spectrum will differ most markedly from the source spectrum
at high photon energies ($E\gtrsim 1$\,MeV)
at which it (the emergent spectrum) will be heavily depleted.
In other words, the observed (emergent) spectrum becomes softer.
Then, the $e^{-}e^{+}$ pairs eventually
annihilate to produce two (infrequently 3) photons,
but usually not one high- and one low-energy photon.

The result is that high-energy photons are preferentially removed from the
observed spectrum. The observation of a measurable amount of quanta with
$E > E_{th}= \sqrt {E_1 E_2}$
is not expected unless the optical depth $\tau_{e^-e^+}$
to pair production is equal to 1 or less,
because the threshold for electron-positron pair production (1)
is also a sensitive function of the angular distribution of the
radiation field in the very source (see below).
Thus, the observation of a considerable number of quanta with $E > 1$\,MeV
due to the filter effect (1) is not expected, if only the optical depth for
the $e^{-}e^{+}$ pair production is not proved $\lesssim 1$ indeed
for various reasons, for example, because of anisotropy of the radiation 
field in the GRB source itself.

   As is seen from the paper by Carrigan and Katz [2],
in 1992 it was generally
accepted that typical energies of most photons in observed GRB spectra
are still rather small. Further in the peper, Carrigan and Katz adduce
the estimates of distances to burst sources of such photons
with the {\it semirelativistic} energies.
The matter is that the problem of a compact source
(in relation to the 1979 March 5 event in LMC)
and a surprisingly big distance arises indeed,
but not because of a problem with the release of
``heavy"\, (100\,MeV, 1\,GeV, or more)
ultrarelativistic photons which interfere with ``light" ($\lesssim 1$\,MeV)
target photons.
%observed in the GRB spectra.
The powerful 1979 March 5 event in LMC was observed without any
super heavy photons in its spectrum.
To make sure of it
one should just look at the spectra of this
burst published by Mazets and Golenetskii in their review [9].

To explain why the effect of the photon ``$e^{-}e^{+}$ confinement"\,
  does not function in
this GRB source (1979 March 5 event in LMC) different possibilities
were discussed [1, 2]. In particular, the authors immediately point
out to the angle dependence (1) of the threshold
of the $e^{-}e^{+}$ production.
A possible ``loophole"\, exists if the source produces a {\it strongly
collimated} beam of photons.
Thus, {\bf the question is about an asymmetry of the radiation field
in the source.}
In this case, even high-energy photons are below the
threshold for the pair production if $\theta_{12}$ is small enough.
The presence of such a ``window"\, in the opacity for {\it collimated} photons
suggests that in a region opaque to pair production
much of the radiation may emerge through this window,
in analogy to the great contribution of windows in the material
opacity to radiation flow in the usual (Rosseland mean) approximation.

The use of the words ``strongly collimated"\, in the (``old") paper [2]
could be somewhat confusing. What means {\it strongly} indeed?
At that time there were no observations of GRB spectra in the region
of high energy $E$.
Heavier photons with $E \sim 10$\,MeV (beyond the peak of $\sim 1$\,MeV)
have been reliably observed only with EGRET/BATSE.
In particular, from formula (1) for such photons an estimation of the
collimation angle can be obtained (without any ``target-photons"):
$1 - cos\theta_{12} = 0.522245\,$MeV$^2/(10$MeV$\cdot 10$MeV$)\approx 0.005.$
It corresponds to $\theta_{12}$ less than $6^o$ only.
It means that
the quanta with energy $\sim 10$\,MeV leaving the source within
a cone of $\sim 6^o$ opening angle
do not give rise to pairs, and all {\it softer} radiation
can be uncollimated at all. So the collision of 10\,MeV quanta
with quanta of lower energy occurs at angles greater than $60^o$
($0.522245\,$MeV$^2/(10\,$MeV$ \cdot 100\,$KeV$) \approx 0.5$), and softer
quanta leaving the source within the cone of such opening angle do not prevent
neither heavy nor (especially) light quanta to go freely to infinity.

Thus, formula (1) demands more or less strong collimation only for {\it
a small part} of the heaviest quanta radiated by the source. If one looks at
energetic spectra of typical GRBs (the same reference to Mazets and
Golenetskii [9]) presented in
%the old way of
$F($cm$^{-2}$s$^{-1}$KeV$^{-1})$~vs.~$E$(KeV) --- {\it the number} of photons
per a time unit in an energy range unit per an area unit versus the photons
energy, --- then everything becomes clear. Only a small part or a small
{\it amount} of quanta/photons observed beyond a threshold of
$\approx 700$\,KeV can be collimated,
but within a cone of $< 90^o$ opening angle.

At present, 6 degrees for 10\,MeV quanta would not be considered
as a strongly collimated beam. Now such opening angles (of\,\, jets)
are considered to be quite suitable in the ``standard"\,  or the most
popular theory of fireballs.
If one proceeds
right away from an idea that it is necessary to release
quanta with the energy up to 10\,MeV, then we
would obtain at once
a version of a collimated theory with the $\Gamma$ of $\sim 10$.
But such a way in the standard fireball theory is a dead end also.
The allowing for an initial collimation of GRB radiation
can drastically change
this model (see below) for the collimation arising
{\it directly in the source}
but not because of a huge $\Gamma$ of $\sim 1000$
what would be needed to solve the compactness problem
if a ultra-relativistic jet is a GRB source indeed.

One way or another, the light flux is to lead to corresponding effects of
radiation pressure upon the matter surrounding the source.
And if in addition the radiation is collimated, then
the arising of jets (at so enormous light flux)
becomes an inevitable consequence of even a small
asymmetry of the radiation field {\it in the source}.

But the question is if:

\section{Is the jet a GRB source or not?}

Indeed, perhaps one should take into account right away this angular
dependence of the threshold of the
pair $e^{-}e^{+}$ production (1) before
the ultra relativistic limit, allowing for a possibility of a preferential
(most probably by a magnetic field) direction in the burst source on the
surface of a compact object -- the GRB source.
Can we do without the radiating and accelerated
jet (in the model of fireball) up to a huge value of the Lorentz factor ,
but supposing that the source of GRB radiation is {\it already} collimated
by the burst source itself (in a compact GRB model)?

The rather strong collimation of GRB \g-rays,
reaching near-earth detectors, can be observably justified if,
due to further accumulation of observational data about coincidence of 
GRBs and supernovae (SNe), it will turn out indeed that
the GRBs could be the beginning of explosions
of {\it usual} massive or core-collapse SNe [24].
%(Sokolov 2001a, 2001b).
At least,
all results of photometrical and spectral observations of GRB host galaxies
confirm the relation between GRB and evolution of a {\it massive star}, i.e.,
the close connection between GRB and
relativistic collapse with SN explosion in the end of the star evolution
[25, 26, 27].
%(Djorgovski et al. 2001; Frail et al. 2002; Sokolov et al. 2001; etc.)
The main conclusion resulting from the investigation of these galaxies is
that the GRB hosts do not differ in anything from other galaxies
with close value of redshifts $z$:
neither in colors, nor in spectra, the massive star-forming rates [27],
and the metallicities [28].
%%%(Savaglio 2006).
It means that these are generally starforming galaxies (``ordinary"\,
for their redshifts) constituting the base of all deep surveys.
In point of fact, this is the first result of
{\it the GRB optical identification} with
already known objects: GRBs are identified with ordinary
(or the most numerous in the Universe at any $z$)
galaxies up to $\approx 26$ stellar magnitudes.
So, with allowing for the results of direct optical identifications
this makes it possible to estimate directly from observations
an average yearly rate of GRB events in every such galaxy
by accounts of these galaxies for the number of galaxies brighter than
26th st. magn. It turns out to be equal to
       $ N_{GRB} \sim {\bf 10^{-8}} yr^{-1} galaxy^{-1}$.
(But most probably this is only an upper estimate [24].)
%, see in Sokolov 2001b).

Allowing for the yearly rate of (massive) SN explosions
       $ N_{SN} \sim 10^{-3} - 10^{-2} yr^{-1} galaxy^{-1}$,
the ratio of the number of GRBs, related with the collapse of massive stars
(core-collapse SNe), to the number of such SNe is close to
$ N_{GRB}/N_{SN} \sim 10^{-5} - 10^{-6}$.
(This is also can be only the upper estimate for Ib/c type SNe [24].)
%(Sokolov 2001a).
Certainly, only the further increasing of the number of coincidences of GRBs
and SNe ({\it identifications of GRBs with Type Ib/c SNe})
should finally tell us whether we have a core-collapse SN
(spanning a large range of luminosities)
{\it in each} GRB or whether the collapse
of a massive star evolves following different paths according to the value
of parameters as mass, angular momentum, and metallicity [29].
But here we proceed from the simplest assumption, which has been confirmed
from 1998
by increasing number of observational facts, that {\it all} long-duration
GRBs are related to explosions of massive SNe.
Then the ratio $ N_{GRB}/N_{SN}$
should be interpreted as a very strict ``$\gamma$-ray beaming"\,
for a part quanta
{\it reaching an observer}, when gamma-ray radiation (a part of it) of the
GRB source propagates to very long distances within a very small solid angle

\begin{equation}
     \Omega_{beam} = N_{GRB}/N_{SN} \sim (10^{-5} - 10^{-6})\cdot 4\pi.
\end{equation}

Another possible interpretation of the so small value of
$ N_{GRB}/N_{SN}$ --- a relation to a rare class of some
peculiar SNe --- seems to be less possible (or hardly probable),
since then GRBs would be related only to the
$10^{-5}-10^{-6}$th part of all observed SNe in distant galaxies
(up to 28th mag).
These are already not simple peculiar SNe, with which the Paczy\'nski's
hypernova is sometimes identified [30].
%(Paczy\'nski 1999)
%Fields et al. 2002).
The peculiar SNe (hypernovae), such as 1997ef, 1998bw, 2002ap,
turn out to be too numerous [31].
%Podsiadlowski et al. 2004)
%Richardson et al. 2002;
On the other hand,
the more numerous are GRB/SN coincidences [29] of type of
GRB\,030329/SN 2003dh, GRB\,060218/SN\,2006aj, or GRB/``red shoulder"\,
in light curves, the more confident will be the idea that GRB radiation
is collimated, but not related to a special class of SNe.
The more so, that explosion geometry features
 (SN explosion can be axially symmetrical)
make the attempts to select a class of ``hypernovae"\, more complex
%(Willingale et al. 2004,
[33] (see the end of their text).
Now there are already other papers [32],
%2003c (Lamb et al. % Donaghy \& Graziani, 2003a, 2003b, 2003c),
pointing out to a possibility of collimated radiation
from the GRB source (2).

Let us suppose that only {\bf the most collimated part}
of gamma radiation get to an observer,
say, along a rotation axis of the collapsing core of a star with
magnetic field.
And if GRBs are so highly collimated,
radiating only into a small fraction of the sky,
then the energy of each event  $E_{beam}$
must be much reduced, by several orders of magnitude
in comparison at least with a (so called)
``isotropic equivalent"\, $E_{iso}$, of a total GRB energy release
  ($E_{iso} \sim 10^{51} - 10^{52}$\,erg and up to $\sim 10^{53}$\,erg):
\begin{equation}
     E_{beam} = E_{iso} \Omega_{beam}/4\pi \sim 10^{45} - 10^{47}\,erg .
\end{equation}

If it is just this case which is realized, and if the energy (3) of \g-rays
propagating in the form of a narrow beam reaching an observer on Earth
is only {\it a part} of the total radiated energy of the GRB source,
then the other part
(from $\sim 10^{47}$\,erg  to $\sim 10^{49}$\,erg, see below)
of its energy can be radiated in {\it isotropic} or almost isotropic way
indeed.
But at the spherical luminosity corresponding to
a total GRB energy of,
e.g., $\sim 10^{45} - 10^{47}$\,erg,
no BATSE gamma-ray monitor detector,
even the most sensitive one, would detect flux, corresponding to so low
luminosity for objects at cosmological distances of $z \gtrsim 1$, and if the
observer is outside the cone of the collimated component of radiation (2).
I.e. (3) can be close to the lower estimate of the total radiated energy
of GRB sources, corresponding to the flux measured within the solid angle
(2), in which the most collimated component of the source radiation is
propagating.
(We always suppose that {\it all} long-duration GRBs
are related to SNe.)
So, there is a possibility at least to considerably {\it reduce} at once
the total (bolometric) energy of GRB explosions.

Apparently, this question
(what radiate: a central compact source or an extent jet?)
is crucial for any GRB mechanism.
If the GRB source radiation (mainly a hard component of
the GRB spectrum) is collimated indeed,
then we will have to return to the old idea:
the radiation (GRB) arises
{\it on a surface}
of a compact object of the order of tens of kilometers(?).
Further we will try to do without an (a priori) assumption that it is only the
jet's ``end"\, which radiates.
The jet arises for sure, but
because of the strong pressure of the collimated radiation on the matter
surrounding a compact (down to $10^7$\,cm and less) GRB source. Certainly,
this jet accelerated by photons up to relativistic velocities will radiate
also, but it would be already an afterglow, but not GRB itself.

\section{The radiation pressure and origin of the jet}

If the scenario: {\it massive star} ---$>$ {\it WR star} ---$>$
{\it pre-SN = pre-GRB} ---$>$ {\it the collapse of a massive star core}
with formation of a shell around WR is true,
then it could be supposed that the reason for arising
of a relativistic jet is the powerful light pressure of the collimated or
non-isotropic prompt radiation of the GRB source onto the matter
of the WR star envelope located immediately around the source
itself --- a collapsing core of this star.

For example, the
radiation field arising around the compact source can be
non-isotropic --- axially symmetric due to magnetic field and effects of
angular dependence (1) of the threshold of the
$e^{-}e^{+}$ pair production.
And only a part ($\sim 10\%$ or
even
$1\%$) of the total GRB energy ($\sim 10^{47} - 10^{49}$\,erg)
may be the collimated radiation within the solid angle (2),
which breaks through the dense envelope
surrounding the collapsing core of the WR star and reaches the Earth.
The main things now are: 1) {\it the collimated flux} of radiation
from the source and
2) existence of {\it dense} gas (windy) environment
pressed up by radiation from the GRB compact source embedded in it.
This environment can be the most dense just near the source,
if the density is close to $n = A r^{-2}$ (the WR law for stellar wind).
Here the distance $r$ is measured from the WR star itself, and
$A\sim 10^{34}$\,cm$^{-1}$ [34].
%{\bf (Ramirez-Ruiz et al. 2001)}.

For the force of light pressure that can act on gas environment
(plasma) around the GRB source (the WR star) we have
$L_{GRB} \cdot (4\pi r^{2})^{-1} \cdot (\sigma_{T}/c)$, where $L_{GRB}$ is a
so called isotropic {\it luminosity} equivalent of the source
($\sim 10^{50-51}$\,erg$\cdot$s$^{-1}$ and more),
$r$ is a distance from the center (or from the source),
$\sigma_{T} = 0.66 \cdot 10^{-24}$\,cm$^2$ is the Thomson cross-section,
$c$ is the velocity of light.
It is clear even without detailed calculation that near the WR core
($r\sim 10^9$\,cm) such a force can over and over exceed (by 12-13 orders)
the light pressure force corresponding to {\it the Eddington limit}
of luminosity ($\sim 10^{38}$\,erg$\cdot$s$^{-1}$ for 1\,$M_\odot$).
The isotropic radiation with so huge luminosity
$L_{GRB}\sim 10^{50-51}$\,erg$\cdot$s$^{-1}$
(or the light pressure)
can also lead to fast acceleration
(similar to an explosion) of environment adjacent to the source.
But if we assume that the radiation of the GRB source is non-isotropic and a
part of it is collimated or we have very strong beaming with the solid angle
   $\Omega_{beam} \sim (10^{-5} - 10^{-6})\cdot 4\pi$,
then the forming of directed motion of relativistic/ultra-relativistic
jets becomes inevitable, only because of so huge/enormous light pressure
affecting the {\it dense} gas environment in the immediate vicinity of the
source - collapsing stellar core.

We can estimate the size of the region {\it within} which such a jet can be
accelerated by the radiation pressure up to relativistic velocities: \\
1.~~~If the photon flux producing the radiation
pressure accelerating the matter at a distance $r$
from the center (near the GRB site) is equal to
   $L_{GRB}\cdot (4\pi r^{2})^{-1}$,
then in the immediate vicinity from the GRB source
(the collapsing core of WR star)
such a flux can be enormous.
It is {\it inside} this region where the jet originates
and undergoes acceleration up to
ultra relativistic velocities. \\
2.~~To accelerate the matter up to velocity of at least $\sim 0.3c$,
at the {\it outer} boundary of this region the photon flux must be at least not
less than the Eddington flux
$L_{Edd}\cdot (4\pi R_{*}^{2})^{-1}$.
Here $L_{Edd}$ is the Eddington limit
$\sim 10^{38}$erg$\cdot$s$^{-1}$  for 1\,$M_\odot$  and $R_{*}$
is the size of a compact object of
$\sim 10^{6}$\,cm.
 (By definition: $L_{Edd}\cdot (4\pi R_{*}^{2})^{-1}$
is a flux {\it stopping } the accretion onto a compact source
--- the falling of matter on the source at a parabolic velocity.
For a neutron star it is equal to $\sim 0.3c$.)

From the condition that the photon flux
$L_{GRB}\cdot (4\pi r^{2})^{-1}$ at distance $r$ is equal to
$L_{Edd}\cdot (4\pi R_{*}^{2})^{-1}$
(or at least not less than this flux),
and taking into account that the luminosity or rather its {\it isotropic
equivalent} of the GRB radiation is
$L_{GRB}\sim 10^{50-51}$erg$\cdot$s$^{-1}$,
it is possible to obtain an estimate of the size of
$\sim 10^{12}$ cm $\approx 14R_\odot$.
At least, at this outer boundary the light pressure is still able to
accelerate the initially stable matter up to sub-light velocities $\sim
0.3c$.
And {\it deeper}, at less distances than $\sim 10^{12}$\,cm from the
source, say, at $r\sim 10^{9}$\,cm (somewhere {\it inside} the region of the
size less than the characteristic size of collapsing core of the massive
star) the light accelerates the matter up to ultra relativistic velocities
with the Lorentz factor of  $\sim 10$ at
$L_{GRB}\sim 10^{50}$erg$\cdot$s$^{-1}$.
It can occur in a rather small volume of the typical size of
  $\lesssim R_\odot$.
Thus, inside the region of a size of less (in any case) than $10-15 R_\odot$,
a relativistic jet arises as a result of the strong light pressure onto the
ambient medium.

\section{Concluding remarks: the observational consequences}

{\it The superluminal radio components:}
From the above-said it follows
that the suggested compact GRB scenario allows also
predicting the behavior of superluminal radio components which, e.g., have
been observed for GRB 030329 [37].
%(Taylor et al. 2004)).
If it is no considerable deceleration of the jet (bullet)
with the Lorentz factor of order 10,
hence we expect that the
superluminal radio components related to the jet have the following
properties: \\
1) the radio component will move with the constant observed superluminal
velocity; \\
2) the characteristic
observed velocity of the superluminal component is of the order of the Lorentz
factor, i.e. of order $10\,c$.

Thus, it is undoubtedly that the GRB radiation is to be collimated
in the compact model with GRB source
  $\sim 10^{8}-10^{6}$\,cm, but the
collimation (2) concerns mainly only a small part of hard quanta.
The pairs production threshold (1) for such quanta naturally and smoothly,
according to the law $(1-cos\theta)^{1/2}$, rises
with the decreasing of the angle $\theta$ between the direction at which the
photon is radiated from the surface of the compact object and
{\it a selected} direction (e.g. the magnetic field) on the surface.
As a result, beside a soft component,
the more and more hard part of the burst spectrum is passing through,
and it is possible to suggest non-isotropic (axially symmetrical)
field of radiation around the source.

{\it The non-collimated XRFs and SNe:}
E.g. the XRFs can be not collimated at all or slightly collimated (XRR GRB),
but with the low total bolometric energy of $\sim 10^{47}$\,erg.
Since  most probably these are actually the explosions of massive SNe
at distances of 100\,Mpc [35, 36],
%(Norris 2003; Norris \& Bonnel 2003),
they can be observed much more frequently than it is predicted by the standard
fireball GRB model.
One should try to find early spectral and photometrical SN features.
Then, in general,
the observational problem of XRF/XRR/GRB identification becomes a
special section in the study of cosmological SNe.
(It will be recalled that the GRB\,030329/SN\,2003dh was a XRR GRB but not
a classical GRB and XRF/GRB\,060218/SN\,2008aj was the X-ray flash [29, 32].)

As to normal classical GRBs and especially those ones with many heavy
quanta in spectra, it is possible to obtain directly from formula (1) a
kinematical estimate of the limit collimation of this $\gamma$-radiation,
which, in turn, independently agrees with the observational ratio
(2) of the yearly rates $N_{GRB}/N_{SN} \sim 10^{-5} - 10^{-6}$.
If the matter concerns quanta with $E\sim 100$\,MeV [7] of distant and the most
distant GRBs, then from
$1 - cos\theta_{12}\approx 0.5$\,MeV$^2/(100$MeV$\cdot100$MeV$) = 0.5\cdot 10^{-4}$
it follows that the radiation of such GRBs turns out to be the most
collimated.
Such photons must be radiated in the cone of an opening of $\approx 0.5^o$
and be detected in the spectra of the rather distant GRBs with $z\sim 1$ and
 farther because of geometrical factor only.

{\it The Amati law:}
Thus, a natural consequence of our compact model of the GRB source
is the fact
that distant bursts ($z\gtrsim 1$)
turn out to be harder ones, while close ``GRBs" ($z \sim 0.1$)
look like XRF and XRR GRBs with predominance of soft X-ray quanta
in their spectra (though the factor $1+z$ also works).
Naturally, the effects of observational selection due to finite sensitivity
of GRB detectors should be also taken into account.
For example, the soft spectral component of the distant (classical) GRBs is
``cut"\, out by the detector sensitivity threshold.
And the isotropic X-ray burst, simultaneous with the GRB, can be simply not
seen in distant (classical) GRBs because of
the  low total/bolometric luminosity of the source
in the compact GRB model ($< 10^{49}$\,erg).
Actually, XRF and XRR GRBs have lower values of
$E_{iso}$ (so called isotropic equivalent),
than GRBs [16, 32].
%(Amati et al. 2002; Lamb et al. 2003a, 2003b, 2003c).
It is an important {\it observational}
result of BeppoSAX and HETE-2 missions.
We mean the detection of obvious XRFs and XRR GRBs first
by BeppoSAX [16]
%(Amati et al. 2002)
and then by HETE-2.
In our compact model of GRB source it (the Amati law)
can be a ``simple"\, consequence of formula (1) + collimation (most
probably) by magnetic field on the surface of the compact object.

{\it The yearly rate of core-collapse SNe and Fast X-ray Trasients:}
In the scenario of jet formation,
which was discussed in this paper
an isotropic X-ray, optical and radio emission of
GRB {\it afterglow} is possible.
At that an initial assumption was just a possibility of the GRB
collimation (2), which follows from the comparison of the rates of
GRBs and SN explosions in distant galaxies.
It means that the close relation between GRBs and SNe
was taken as the basic assumption:
{\it all} long GRBs are always accompanied by SN explosions, which are
sometimes observed, and sometimes not [24].
%(Sokolov 2001a, 2002).
In other words, the long GRB is {\it the beginning} of a massive
star collapse
or the beginning of SN explosion, and GRBs must always be accompanied
by SN explosions (of Ib/c type or of {\it other} types of massive SNe).
Then in any case the total energy release at the burst in \g-rays can be
{\it not more} than the total energy released by any SN ($ <$ or $\sim
10^{49}$\,erg) in all {\it electromagnetic} waves.
But with so ``low"\, total energy of the GRB explosion
($\lesssim 10^{49}$\,erg) the only possibility
to see GRB at cosmological distances ($z\gtrsim 1$)
is the detection of at least the most collimated part
of this energy ($1-10$\%) leaving the source within the solid angle of
$\Omega_{beam} \sim (10^{-5} - 10^{-6})\cdot 4\pi$.
The rest can be inaccessible for GRB {\it detectors} with
a limit sensitivity of $\sim 10^{-7}$\,erg$ \cdot $s$^{-1}\cdot $cm$^{-2}$.
Certainly, it does not concern the 10~000 times more sensitive X-ray
{\it telescopes} which were used to make sky surveys with the
Ariel\,V, HEAO-1, Einstein satellites [15].
%(Heise et al. 2001).
For the limit sensitivity of
$\sim 10^{-11}$\,erg$\cdot $s$^{-1}\cdot $cm$^{-2}$ in
the band of $0.2-3.5$\,keV the X-ray observatory
(Einstein) recorded {\it Fast X-ray Trasients}
(unidentified with anything) at a rate of $\sim 10^6 yr^{-1}$
all over the sky.
It agrees well with an average rate of the massive SNe explosions
in distant galaxies,
but for the present, GRB-detectors see only $\sim 10^{-4}$
part of this huge
number of the distant SN explosions as GRBs.

It is natural that at the total/bolometric energy of ``GRB"\,
$\sim 10^{47} - 10^{49}$\,erg and at the GRB energy (3) released in
the narrow cone (2), ``the fireball"\, also looks in quite a different way.
As to the compactness problem solved by the fireball model
for GRB energies of $10^{52} - 10^{53}$\,erg, there is no such a
problem for ``\g-burst"\, energies $\sim 10^{47} - 10^{49}$\,erg.
In any case, allowing for the low \g-ray collimation from the surface
of the compact object --
GRB/XRR/XRF source, which is necessary for the
angular dependence of $e^-e^+$ pair production (1), this problem is
solved under quite different physical conditions
in the GRB-source than that supposed by Piran [18].
In the scenario:
{\it massive star} ---$>$ {\it WR} ---$>$ {\it pre-SN $=$ pre-GRB},
in which only a small part of the most collimated radiation with the
collimation (2) goes to infinity and, correspondingly, with the
total energy of $10^4 - 10^6$ times less than in the standard theory, the
source can actually be of a size $\lesssim 10^8$\,cm.
It means that at the energies of up to $\sim 10^{49}$\,erg the old
(``naive") estimate of the source size resulting directly from the time
variability of GRB can be quite true.
Thus, the point can be that the burst energy is much less than in the
standard fireball model.

{\it The strong polarization of the GRB radiation:}
But in such a model [38] the compact source must always have some radiating 
{\it surface} (but not an event horizon) and, respectively, always occupy
some finite volume.
Such an object can have both a strong regular magnetic field and
a nonuniformly-raidating surface connected with it.
The radiation field arising around the source could be anisotropic,
e.g., axially symmetric due to the local magnetic field.
In particular, non-uniform radiation at the source surface (e.g. polar caps)
could lead to efficient collimation or anisotropy of the radiation field [2],
due to the influence of the angular dependence (1) for the
e- e+ pair-creation threshold.
Such anisotropy could be associated with the transport of radiation
in a medium with a strong ($\sim 10^{14}-10^{16}$\,G) magnetic field ,
when the absorption coefficient for photons polarized orthogonal
to the magnetic field is very small [39, 40].
In this case, the observation of strong linear polarization
of the GRB radiation should be another consequence of our compact GRB model.

We always suppose that {\it all}
long-duration GRBs are related to core-collapse SNe,
or the rate of GRB-SNe is the rate of all massive star deaths.
Thus, in the compact model of GRBs,
the formation of massive ($\gtrsim 3M_\odot$) and compact remnants
of the core-collapse SNe (with the massive progenitor stars
$> 30-40\,M_\odot$) can be always accompanied by the GRB (or XRF) phenomenon.
But the observations should finally tell us whether we have a SN
in each GRB or whether the collapde of a massive star evolves following
different paths. Whatever the answer may be the fundamental point of the
connection is that GRBs may serve as a guideline to better understand
the mechanism, and possibly solve the long standing problem of the
core-collapse SN explosion, since in the GRBs we have additional information
related to the core-collapse [41].

\end{document}